\documentclass[twocolumn,showpacs,preprintnumbers,amsmath,amssymb]{revtex4}
\usepackage{graphicx}
\usepackage{dcolumn}
\usepackage{bm}

\begin{document}

\title{Multiple electronic components and Lifshitz transitions 
by oxygen wires formation in layered cuprates
and nickelates}

\author{Thomas Jarlborg$^{1,2}$ and Antonio Bianconi$^{2,3,4}$}

\affiliation{
$^1$DQMP, University of Geneva, 24 Quai Ernest-Ansermet, CH-1211 Geneva 4,
Switzerland
\\
$^2$ RICMASS Rome International Center for Materials Science Superstripes, Via dei Sabelli 119A, 00185 Rome, Italy
\\
$^3$ Institute of Crystallography, Consiglio Nazionale delle Ricerche, CNR, Monterotondo, Roma, I-00015, Italy
\\
$^4$ National Research Nuclear University MEPhI (Moscow Engineering Physics Institute), 115409 Moscow, Russia }


\begin{abstract}

There is a growing compelling experimental evidence that a quantum complex matter scenario 
made of multiple electronic components, and competing quantum phases is needed
to grab the key physics of high critical temperature (T$_c$) superconductivity in layered cuprates.  
While it is known that defects self-organization controls T$_c$, the mechanism remains an open issue.
Here we focus on the theoretical prediction of the multi-band electronic structure 
and the formation of broken Fermi surfaces generated by the self-organization of oxygen 
interstitials O$_i$  atomic wires in the spacer layers in 
HgBa$_2$CuO$_{4+y}$, 
La$_2$CuO$_{4 \pm \delta}$ and 
La$_2$NiO$_{4 \pm \delta}$, 
by means of self-consistent Linear Muffin-Tin Orbital (LMTO) calculations. 
The electronic structure of a first phase of ordered O$_i$ atomic wires 
and of a second glassy phase  made of disordered O$_i$  impurities 
have been studied through supercell calculations. 
We show the common features of the influence 
of O$_i$ wires in the electronic structure in three type of materials. 
The ordering of O$_i$ into wires lead to a separation 
of the electronic states between the O$_i$ ensemble and the rest of the bulk. 
The  wires formation produce first quantum confined localized states
 near the wire which coexist with second delocalized states
 in the Fermi-surface (FS) of doped cuprates. 
In this new scenario for high $T_c$ superconductivity, Kitaev wires with Majorana bound states 
are proximity-coupled to a 2D d-wave superconductor in cuprates.
 
\end{abstract}


\maketitle

\section{Introduction.}

The mechanism behind the emergence of high  $T_c$ superconductivity remains object 
of high scientific interest \cite{lif-h3s-2,lif-h3s-1,lif-h3s-3}.
All high $T_c$ superconductors show a superconducting dome, 
centered at the maximum  $T_{Cmax}$ tuning the chemical potential
by doping or pressure. 
The  high  $T_c$ superconductivity violates the standard approximations of BCS theory \cite{bcs}: 
1) the $dirty$ $limit$ approximation reducing the electronic structure to a single effective Fermi surface and
2) the $Migdal$ approximation where chemical potential is far away from band edges. 
Considering a rigid lattice and neglecting electronic correlations the density-of-states (DOS) at the Fermi energy ($E_F$) in cuprates is not high,
which is in sharp contrast to what  standard BCS theory  of superconductivity would suggest.
The exchange of Cu with Ni permits a much higher DOS at  $E_F$ without major changes of the lattice, 
but the nickelates are not superconducting. 

The popular theoretical paradigms assuming a rigid tetragonal structure for the CuO$_2$ plane 
and a single electronic component model have been abandoned.
The multiband complexity arising from the competition between the effects 
of multiple orbitals in dopant induced 3d$^9$L(a$_1$)  and 3d$^9$L(b$_1$)  electronic states in the Mott Hubbard charge transfer gap \cite{seino,pom},
misfit strain \cite{agrestini,agrestini2},  incommensurate lattice modulations \cite{bia2}, spin density waves, charge density waves \cite{campi3},  
and dopants nanoscale organization \cite{campi3,campi1} need to be 
clarified to understand the quantum complex matter scenario scenario made of multiple interacting complex networks \cite{campi4,gin}.

Recently the consensus is growing on the proposal that the high $T_c$ superconducting dome  \cite{bia1} in cuprates is confined between
topological Lifshitz transitions  \cite{bia3a} for the appearing of new spots of Fermi surfaces \cite{bia2,lif1,lif2}.
typical of multiband superconductors \cite{multi1,lif_a} and it is predicted by the solution of Bogoliubov equations in the stripes scenario.
The electronic structure is strongly modified  by stripes pattern formation with formation of Fermi arcs. 
determined either by 1D short range polaronic Wigner charge density waves 
and lattice effects as lattice tilts typical of perovskites under strain and bond disproportionation associated with the 3d$^9$L states induced by doping 
and the self-organization of oxygen interstitials and defects are expected to contribute to the topological Lifshitz transitions 
in the quantum complex matter scenario of high temperature superconductors. 

The experimental fact that the critical temperature $T_c$ in cuprates appears be controlled by interstitials and defects self organization  \cite{frat,poccia1,little}   
has motivated us to study the electronic structures of large supercells 
of some cuprates and nickelates containing ordered atomic 'wires' of oxygen ions on
interstitial positions.

The systems that we have considered  here are La$_2$CuO$_{4 \pm \delta}$ (LCO) \cite{lco},  La$_2$NiO$_{4 \pm \delta}$
(LNO) \cite{lno}, and  HgBa$_2$CuO$_{4+y}$ (Hg1201) \cite{hbco}. Self-consistent Linear Muffin-Tin Orbital (LMTO) band calculations for supercells
of different sizes have been made for the structures with and without oxygen  interstitial ($O_i$) impurities, 
The total DOS in the three systems near $E_F$ show large, 
seemingly chaotic, peaks, as the number
of $O_i$ are added, even if they are ordered in "wires". However, by comparing the Fermi
surfaces (FS) in the Brillouin Zone (BZ) for the elementary cell and the supercells for the cuprate systems we concluded that
most of the normal FS of the CuO$_2$ layers remain intact despite the additions of ordered impurities. The large DOS peaks are localized on
the impurities with little spillover to nearest neighbors. A long-range charge transfer of electrons from the CuO (and NiO)
layers on the the $O_i$-wires (p-doping) will modify the size of the typical FS cylinder in the case of cuprates.
In this paper we review some of the results and add some complimentary information in order to get a global picture of wire formation
in these systems.

Defect self-organization has been found to control the critical temperature in $Sr_2$$CuO_{4-y}$  \cite{geba}, in
$Sr_{2-x}$$Ba_x$Cu$O_{3+y}$  \cite{gao},   in ($Cu_{0.75}$$Mo_{0.25}$$Sr_2$Y$Cu_2$$O_{7+y}$  
with $0<y<0.5$  \cite{chma1,chma2},  and in $BaPb_{1-x}Bi_xO_3$ \cite{gallo}.
The oxygen interstitial organization in  oxygen doped cuprates $La_2$Cu$O_{4+y}$ 
has been studied by scanning micro x-ray diffraction    
 \cite{frat,poccia1,ricci1,ricci2,ricci3,ricci4,campi1,campi2},  and by STM  \cite{zel1,zel2,zel3}
showing superconductivity emerging in a nanoscale phase separation with a
complex geometry \cite{superstripes,kresin,gin12,gin12a,gin13}
which is determined by the proximity to a electronic topological Lifshitz transition 
in strongly correlated electronic systems \cite{kugel1,kugel2,kugel3} .
In fact it has been found that the domes of high critical temperature occurs 
by tuning the chemical potential near topological Lifshitz transitions 
 in many different cuprates \cite{lifshitz1,lif-cup-1,lif-cup-2,lif_a,lif_b}
 including the case of pressurized sulfur hydride  \cite{lif-h3s-1,lif-h3s-2,lif-h3s-3}.

A considerable theoretical work has shown how the electronic states near the Fermi level
 respond to the lattice and dopants organization changing the topology of the Fermi 
 surface \cite{tjapl,lco,lno,hbco,jbmb,jbb,tj3,js,tj11}
The interest is focused to the perspective that a 
the quasi one-dimensional ordering of dopants could generate stripes giving
a quasi 1D  electronic structure at the Fermi level. 
The new periodicity driven by oxygen interstitial self organization sets up a potential
modulation, which generates unconventional topological Lifshitz transition for the appearing of new bands.

Doping is crucial for high-$T_c$ superconductivity. 
It is usually controlled by the exchange of
the heavy atoms with different valency, like Sr (or Ba) for La in La$_{2-x}$Sr$_x$CuO$_{4}$.
But varying the O-occupation has also been proven to be efficient for doping, either as a
vacancy or as an interstitial impurity.
The Fermi surface for doping larger than 0.21 holes per Cu sites is expected to be predicted by band structure calculations. 
Superconductivity can be enhanced by ordering of oxygen interstitials in cuprates like
La$_2$CuO$_{4 \pm \delta}$   \cite{poccia1}. 
 Band calculations show that oxygen vacancies in the apical positions in 
Ba$_2$CuO$_{4 - \delta}$ (BCO, with $\delta \approx 1$)) make its electronic structure very similar to that of
 optimally doped La$_2$CuO$_{4}$ (LCO) \cite{jbmb,jbb}.
The $T_c$ of BCO is reported to be much larger than in LCO \cite{gao}. 
Self organization of oxygen interstitials enhances $T_c$ \cite{poccia1}, and the Fermi
surface (FS) can become fragmented by oxygen self organization \cite{lco}.
However, the exact role of $ordering$ of the defects is not well known in many cases.
Recently experimental results have been reported on self organization of oxygen interstitials
 in doped cuprates HgBa$_2$CuO$_{4+y}$   \cite{campi3,campi4}. by scanning micro x-ray diffraction
which provide complementary information on local nanoscale structure investigation using x-ray absorption spectroscopy 
\cite{xaneshg1,xaneshg2,xaneshg3} using  EXAFS and XANES  methods \cite{doniach,garcia1}
which probe the deviation of the local structure from the average structure.
In this work we discuss electronic structure results for the hole-doped oxygen-enriched 
 HgBa$_2$CuO$_{4.167}$, (Hg1201), La$_2$CuO$_{4+\delta}$ (LCO) and La$_2$NiO$_{4+\delta}$ (LNO).
The method of calculation is presented in sect. II,  
in sect. III we discuss the results, and
conclusions are in sect. IV.

\section{Method of calculation.}

The calculations are made using the linear muffin-tin orbital (LMTO) method \cite{lmto,bdj} and the
local spin-density approximation (LSDA) \cite{lsda1,lsda2}. 
The details of the methods have been published earlier 
\cite{bj94,tj7,tj1,tj11,lco,lno,hbco}.

\begin{figure}
\includegraphics[width=0.98\columnwidth]{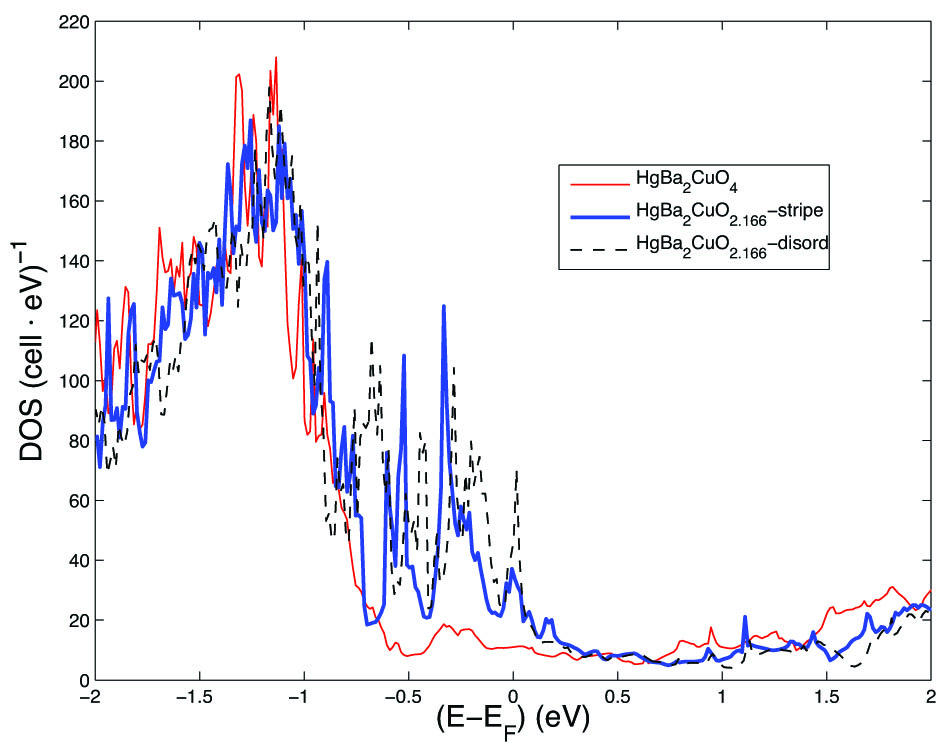}
\caption{(Color online)  The total DOS for Hg$_{12}$Ba$_24$Cu$_{12}$O$_{48+N}$ with $N$=0 (thin red) and $N$=2
(bold blue).
The (black) broken line is when the two O atoms occupy sites far from each other ("disordered"). }
\label{fig1}
\end{figure}


\begin{figure}
\includegraphics[width=0.98\columnwidth]{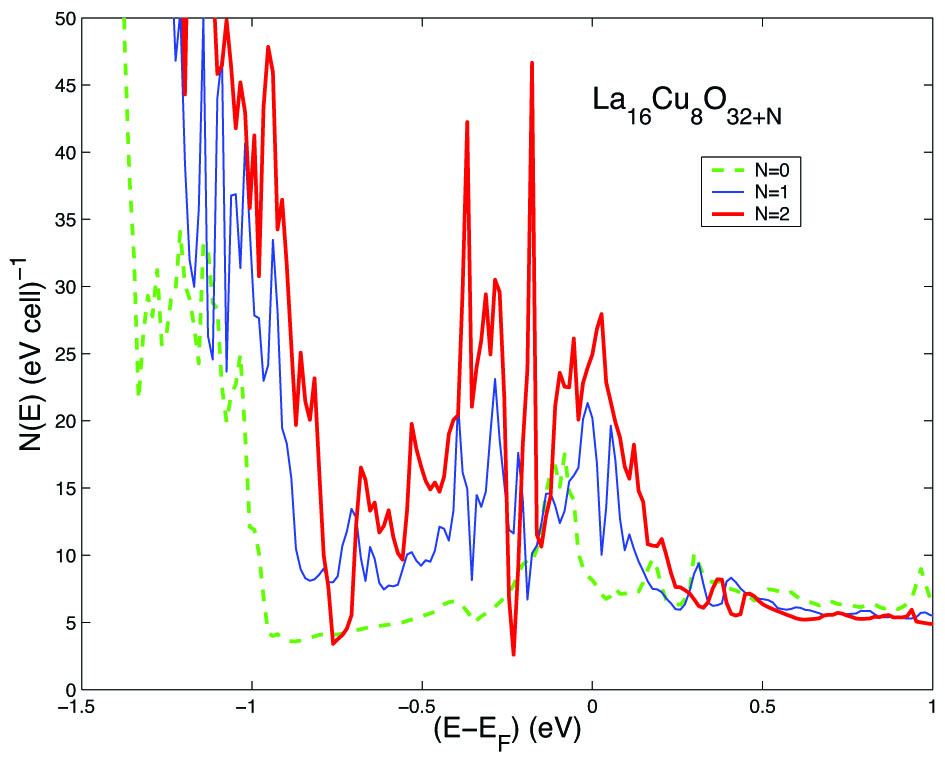}
\caption{(Color online) The total DOS for La$_{16}$Cu$_8$O$_{32+N}$ with $N$=0, 1 and 2 near $E_F$. The Cu-d band
edge moves upwards with increasing N, which is an effect of p-doping. Large DOS peaks near $E_F$ are localized at the $O_i$
impurities. The $O_i$-wire for N=2 becomes electronically isolated from the rest of the p-doped cuprate system, but this cannot
be seen from the DOS functions.}
\label{lcofig}
\end{figure}

\begin{figure}
\includegraphics[width=0.98\columnwidth]{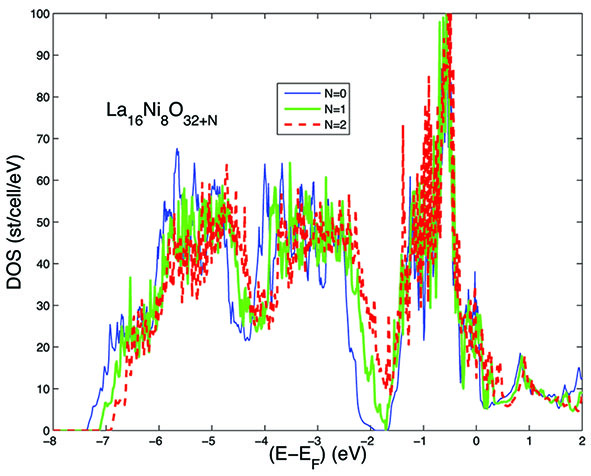}
\caption{(Color online) The total DOS for La$_{16}$Ni$_8$O$_{32+N}$ with $N$=0, 1 and 2. The number of
fully occupied O-p bands in the range -7 to -2 eV increases with increasing N. Nevertheless, a weak effective p-doping
can be detected by the Ni-d band edge as it moves through $E_F$ for increasing N.}
\label{lnofig}
\end{figure}

The supercell Hg$_{12}$Ba$_{24}$Cu$_{12}$O$_{48+2}$ is extended 6- and 2-lattice constants
along $x$ and $y$, respectively. The 2 additional oxygens are inserted in the Hg-plane, as is
known to be the position of excess O in HBCO. These oxygens form a stripe running along $y$.

Calculations for 
the elementary cell of HBCO need 8 atomic sites and 5 "empty spheres", which are included in 
 the most open part of the structure, see ref. \cite{bj94}. The empty sphere in the Hg plane,
 at (0.5, 0.5,0), is the 
location of excess oxygen. 
The lattice constant $a_0$ is 3.87 \AA, and c/a=2.445.
The elementary cell is extended 6$a_0$ along $x$ and 2$a_0$ along $y$. The empty spheres at
(0.5,0.5,0) and (0.5,1.5,0), and 
at (0.5,0.5,0) and (2.5,1.5,0) are occupied by O in the "stripe" supercell and
"disordered" supercells, respectively. This is in the latter case the most distant and uncorrelated
choice for the two interstitial O impurities.  

The elementary cell of La$_2$NiO$_4$ (LNO) contains La sites
at (0,0,$\pm$.721c), Ni at (0,0,0), planar O's at (0.5,0,0) and (0,.5,0) and apical
O's at (0,0,$\pm$.366c), in units of the lattice constant $a_0$=3.86 \AA, where c=1.16.
In addition to the MT-spheres at the atomic sites
we insert MT-spheres at positions (.5,0,$\pm$.5c) and (0,.5,$\pm$.5c) to account for the positions
of empty spheres. The NIO supercells consist of six- and 8-fold repetition of the elementary cell along the plane axis.

In La$_2$CuO$_4$ (LCO) the sites are at the corresponding positions, but with the lattice constant a = 3.8  $\AA$ and c = 12.4 $\AA$  . In addition to the MT spheres at the atomic sites we insert MT spheres at positions (0.5,0,±0.5c) and (0,0.5,±0.5c) to account for the positions of empty spheres.
An elongated supercell is created by an eight-folded repetition of the elementary cell in the diagonal direction of the CuO$_2$ plane.

The doped cases are simulated in calculations where one or two empty interstitial sites have been replaced by O-i atoms.
The total number of sites are different for the different calculations and with different total number of k-points. Information
about this, the size of atomic spheres, basis set and so on can be found in the previous papers \cite{hbco,lco,lno}.
Note that our work is based on density functional theory (DFT) where correlation is a global function of the spin densities \cite{lsda1,lsda2}. 
Strong, non-DFT correlation is not expected for cuprates and nickelates with doping away from half-filling of the d-band. 
ARPES (angular-resolved photoemission
spectroscopy) and ACAR (angular correlation of positron annihilation radiation) on cuprates have detected FS's and
bands in agreement with DFT
calculations \cite{pick,dama,posi}. Here we discuss mainly paramagnetic band results, but the calculations for LNO lead
to ferromagnetic (FM) ground states, even though the FM moments decreases on Ni sites near to the O-i wires \cite{lno}.


\section{Results.}

The total DOS functions at the Fermi level for three supercells of the doped Hg-based cuprate are 
shown  in Figs. \ref{fig1}. The undoped DOS agree with the DOS calculated previously
for one unit cell of HBCO using also $\ell=3$ states for the Ba sites \cite{bj94}. 
A difference is that the less dense
k-point mesh for the supercell makes the DOS curve less smooth. The total DOS at $E_F$ per elementary cell 
is about 1.0 $(eV)^{-1}$ compared to 0.92 here for the supercell.
A calculation of a supercell of intermediate size (Hg$_{4}$Ba$_{8}$Cu$_{4}$O$_{17}$)
corresponding to an impurity concentration of 0.25 shows that the DOS at $E_F$ increases by a factor
of two to about 1.8 per elementary cell \cite{js}. All atoms are close to the impurity in that case,
which explains that the local peaks in the states are not so narrow as in the present case. Here
$E_F$ is on a narrow peak in both cells with O impurities, which makes the DOS higher,
about 3 and 4 $(eV)^{-1}$ per elementary cell for the striped and disordered case respectively.
The increase of the DOS is limited to the first layers of atoms adjacent to the impurity.

\begin{figure}
\includegraphics[width=0.98\columnwidth]{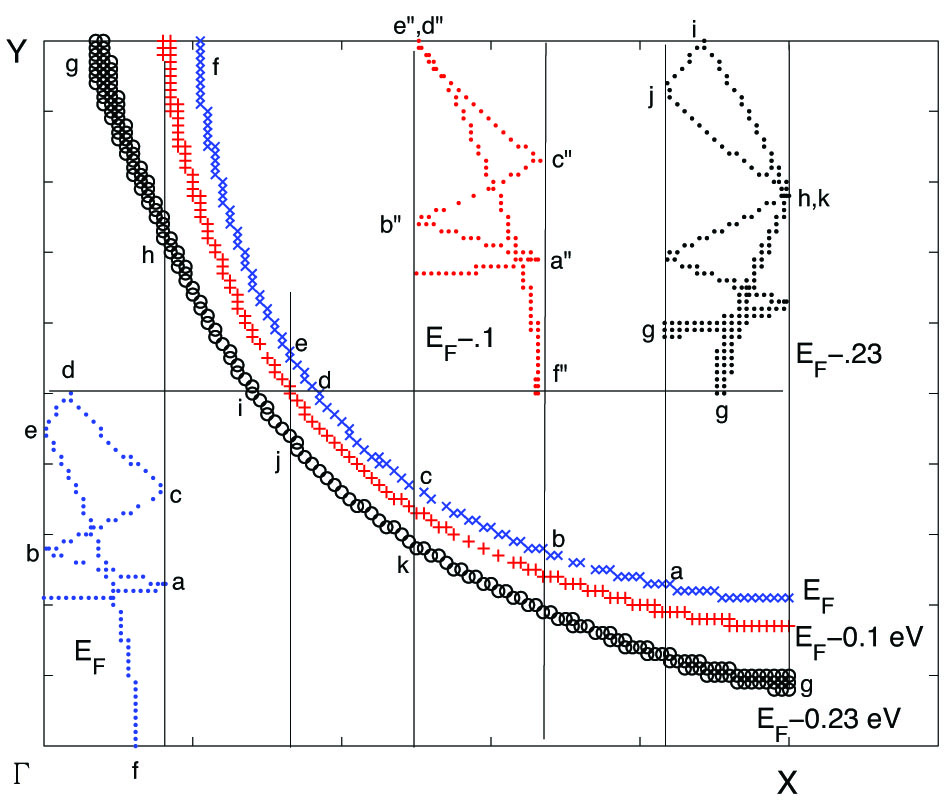}
\caption{(Color online) Upper large panel ($\Gamma$-$X$-$Y$) show the Fermi surface for one unitcell
of undoped HBCO. The "x" is at the calculated $E_F$, the "+" for 0.1 eV
down shifted, and "o" for 0.23 eV down shifted $E_F$. The small panels show how these FS look
after folding the bands into the BZ for the 6x2 supercell.}
\label{fig2}
\end{figure}

\begin{figure}
\includegraphics[width=0.5\columnwidth]{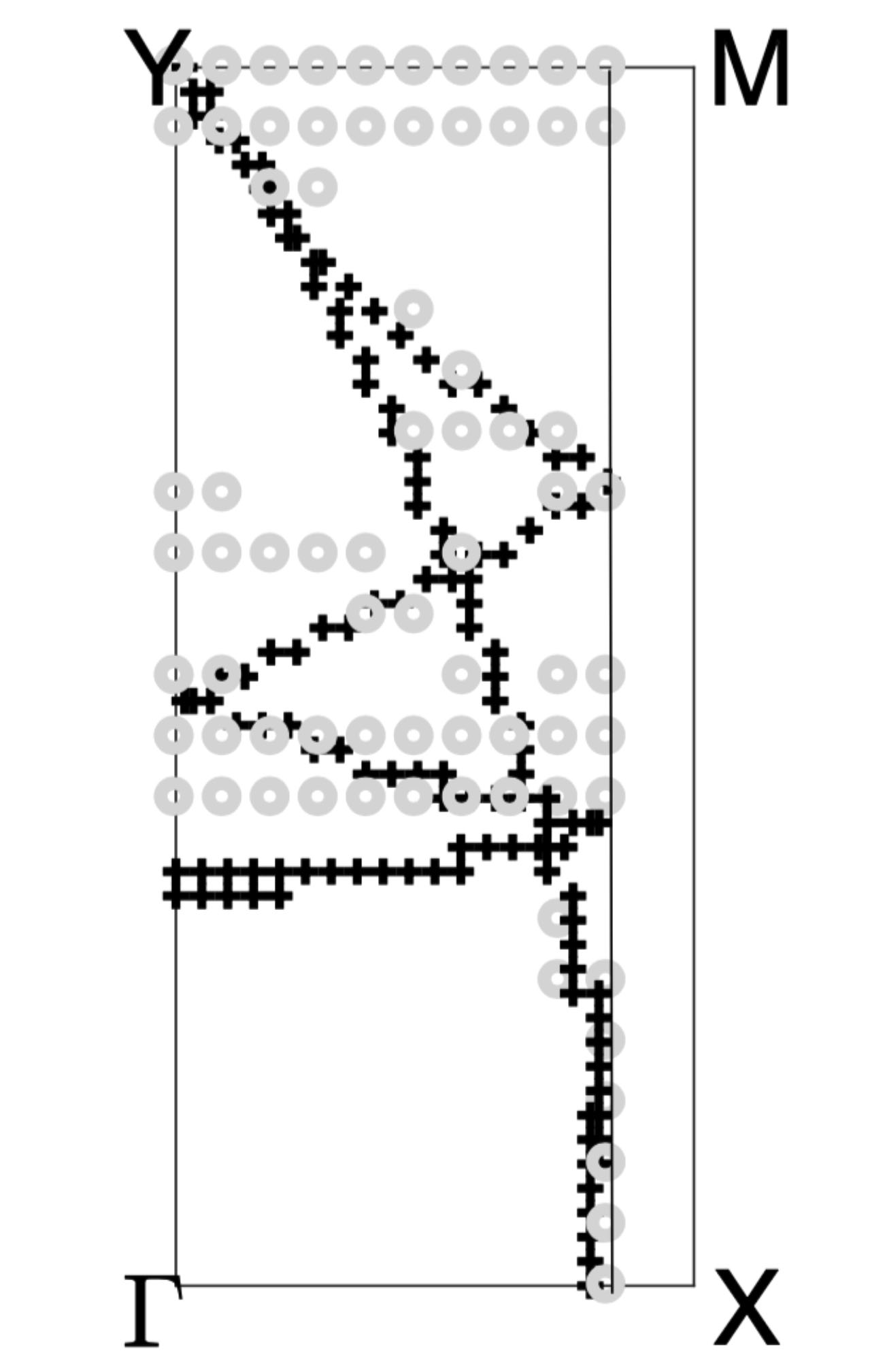}
\caption{(Color online) The FS of Hg$_{12}$Ba$_24$Cu$_{12}$O$_{48+N}$ for N=2 ('O-i wires', gray circles), and
the folded FS for elementary HBCO with $E_F$ down shifted 0.1 eV (black +-signs), i.e. as in the small middle panel in Fig. \ref{fig2}. 
The two FS agree well, which show that the essentials of the FS-cylinder remain even when the O$_2$-wire is present, but
some p-type doping is present since $E_F$ is reduced by 0.1 eV.  The slight deviations for the bands crossing the $\Gamma - Y$ line 
can be understood from an unfolded supercell FS, in which the Fermi radius is shorter towards $X$, 
points $a$ and $b$ (cf. Fig. \ref{fig2}). The radius is also smaller than for undoped HBCO, while near points $c, d$ and $e$ the the radius
is larger and the agreement with undoped HBCO is good.
 However, there is a new band and a new FS at the top of the panel for the supercell. This band has a very high DOS on the O-i sites, and its FS
 cannot be explained by from the simple folding.}
\label{fig3}
\end{figure}

The addition of two ordered or disordered $O_i$ in the cell makes new states to appear near $E_F$ in the
interval -0.6 to +0.2 eV, see Fig. \ref{fig1}. Therefore, it is not possible to judge about the effects of doping
on the FS from these DOS functions. However, it is possible to see that the d-band edge near -1 eV moves to the right, about 0.1 eV,
when two ordered $O_i$ have been added. This is an indication of an effective p-doping, i.e. equivalent of moving
$E_F$ to the left on the standard cuprate DOS. This finding is less evident for the disordered case.
The evolution of the DOS with doping in LCO and LNO can be seen in Figs. \ref{lcofig} and \ref{lnofig}.

In order to reveal the effect of doping on the FS we compare the FS of real calculation for the supercell with that of the
folded FS of the simple FS as calculated for the single unit cell. The FS for one unit cell of HBCO is shown in Fig. \ref{fig2}.
Three circular FS's are shown within one quarter of the BZ, all centered at the upper right corner of the BZ. One FS with the
largest radius is for the calculated $E_F$, and two smaller FS are for assumed larger p-doping when $E_F$ are 0.1 and
0.23 eV lower. Each of the smaller panels of Fig. \ref{fig2} shows the extent of the BZ of the 6*2 supercell, and in three
of them we show how the folded FS's would look like after folding. In the case of intermediate doping, shown by $E_F$-0.1,
one can note that the points marked as $i$ and $j$ in the FS at $E_F$ (or $e$ and $d$ in the FS for $E_F$-0.23) has merged into
one point (marked by $e"$ and $d"$). This merged feature also appear in the FS calculated for the doped cell with ordered $O_i$,
see Fig. \ref{fig3}. The folded FS for $E_F-0.1$ (+-signs) agree well with the real FS for the supercell (o-signs) almost coincides near the
Y-point, which is an evidence of merged branches. Also at the X-point there is a good agreement, which is an indication of a
FS with a radius close to what is indicated by $f$ and $f"$ in Fig. \ref{fig2} for the case $E_F$-0.1. The FS structures
for folded and real FS in the middle of Fig. \ref{fig3} have the same shapes. This shows that a circular FS survives despite
the doping of ordered $O_i$, but that the radius near point $b$ in Fig. \ref{fig2} should be a bit smaller, i.e the circle
seems to slightly retract  near this point.

\begin{figure}
 \includegraphics[width=0.98\columnwidth]{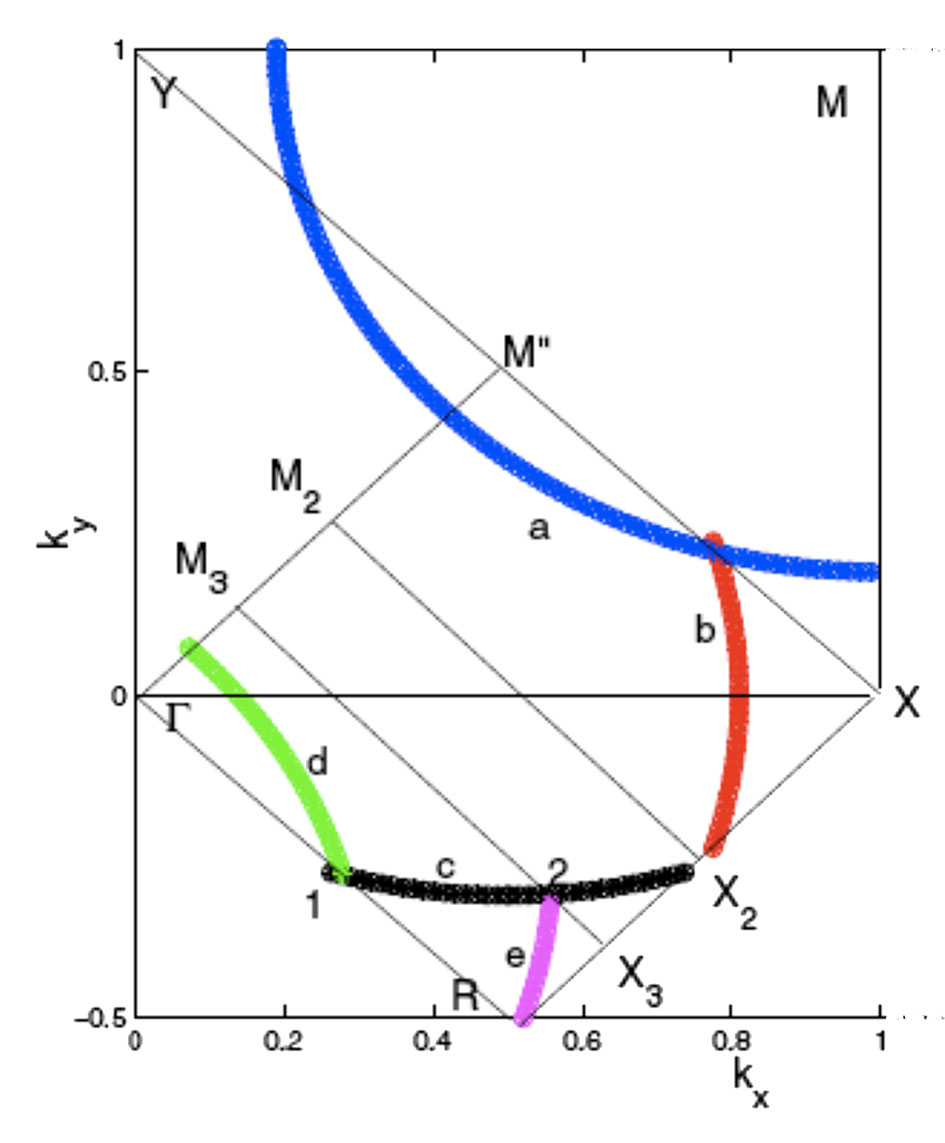}
\caption{(Color online) A schematic view of how the circular FS of doped LCO is downfolded into the
BZ of the LCO-8 supercell. The BZ for one unit cell is limited by $\Gamma$-$X$-$M$-$Y$, and it is
downfolded into $\Gamma$-$R$-$X_3$-$M_3$ for the BZ of the supercell.}
\label{fig4}
\end{figure}

\begin{figure}
\includegraphics[width=0.98\columnwidth]{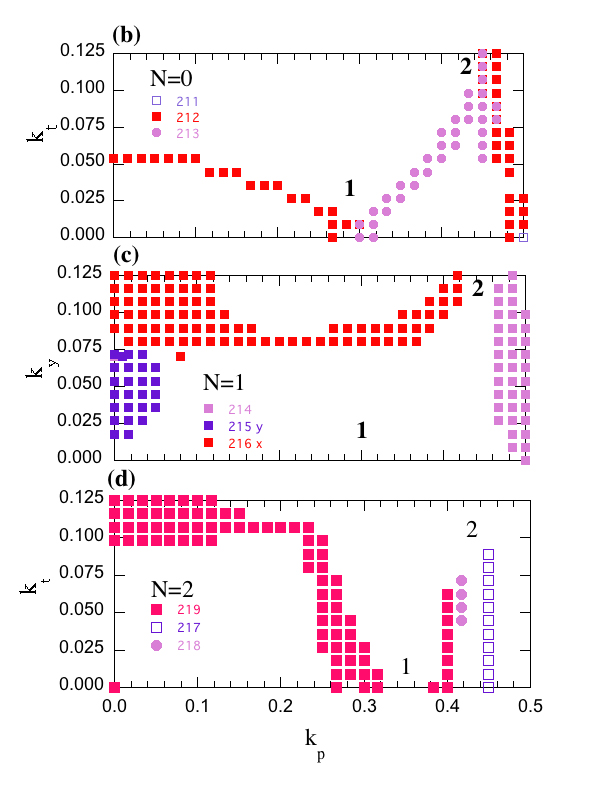}
\caption{(Color online) The three panels show the FS for the 3 calculations with no (N=0, upper panel), one
(N=1, middle) and two (N=2, lower panel) $O_i$ impurities. As expected, the FS for N=0 agree well with the downfolded
FS shown in figure \ref{fig4}. With doping, N=1 and N=2, there is a gradual displacement of the left part of the FS branch,
which corresponds to a larger radius of the unfolded FS (near $M^{'}$), i.e. to a slight p-doping. However, the FS
close to the $X$-point becomes distorted.}
\label{fig5}
\end{figure}

 Thus, there are convincing indications that the typical cuprate FS survives despite ordered $O_i$ doping. However, on the 
 upper edge of the BZ in Fig. \ref{fig3} is seen a clear FS that cannot be understood from folding. A close look at the states
 making this FS branch shows that it consists of $O_i$-p states. The DOS is very high on $O_i$ with only little spill over
 on to the nearest neighbors.
Hybridization with the p-states on the oxygen impurity atoms makes fairly large s- and d-DOS on Hg, and
an increase of the p-DOS on the nearest apical oxygen states, while the influence on planar O and Ba is
not large. The Cu sites are quite distant from the O$_i$, and 
the Cu d-DOS increases about 35 percent near
the impurity, and by 25 percent on the more distant Cu compared to the undoped case. This increased DOS on Cu is similar as
the increased DOS that follows from p-doping, i.e. from a lowering of $E_F$ when it is approaching the
van-Hove DOS peak as in other cuprates. Again, this is a hint that p-type doping is a result of the addition of $O_i$.
When the dopants are disordered we also expect some p-doping effect on the d-band edge (see above), but as is shown in ref.
\cite{hbco} the FS appears to be very distorted compared to the folded FS, and there is no clear separation of states between
$O_i$ and the rest of the lattice sites.
A similar conclusion about doping can be drawn from the results for doped La$_2$CuO$_{4 + \delta}$ (LCO) \cite{lco} despite the
differences of structure and distribution of $O_i$.  The effective p-doping due to increasing $O_i$ content can be
seen from the displacement of the Cu-d band edge, see Fig. \ref{lcofig}. The downfolding is shown in Fig. \ref{fig4}  and \ref{fig5}  as
for HBCO in Fig. \ref{fig2}, but now the folding is fourfold of the AFM double cell (La$_4$Cu$_2$O$_8$)
in the diagonal direction. 

Three branches ($c, d, e$) are seen in the folded BZ limited by $\Gamma$-$R$-$X_3$-$M_3$ when
the circular FS is downfolded. The top panel of Fig. \ref{fig5} show the FS of the undoped supercell (La$_{16}$Cu$_8$O$_{32}$).
The three FS branches from the bands 211-213 agree well with what is expected from the folding. With one $O_i$ (N=1), which makes
an ordered repetition of quite distant impurities, and thus no "wire", one obtains a quite distorted FS. With two $O_i$ (N=2)
and a more wire-like order, one retrieves more of the standard FS in the left part of the BZ, but it has moved upwards compared
to the case N=0, which can be interpreted as coming from a larger radius of the unfolded FS and p-doping. The bands in the right
part of the BZ can be traced to the original FS circle, but gaps have appeared. 

\begin{figure}
\includegraphics[width=0.95\columnwidth]{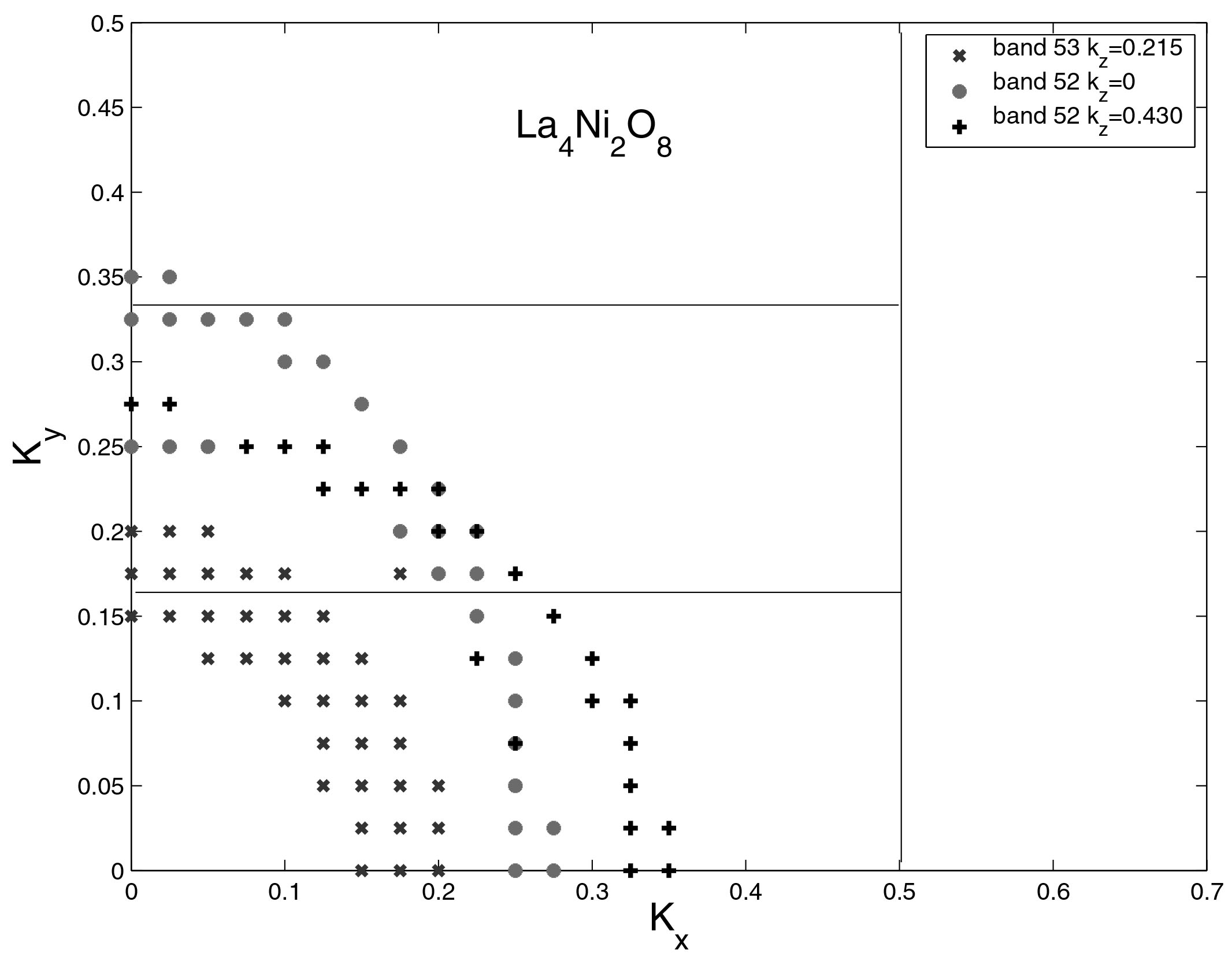}
\caption{(Color online) The FS for the double cell La$_4$Ni$_2$O$_8$, in which three panels show how the BZ is to be folded
to correspond to the BZ for the threefold cell, La$_{12}$Ni$_6$O$_{24}$, see Fig. \ref{fig7}.}
\label{fig6}
\end{figure}

\begin{figure}
\includegraphics[width=0.95\columnwidth]{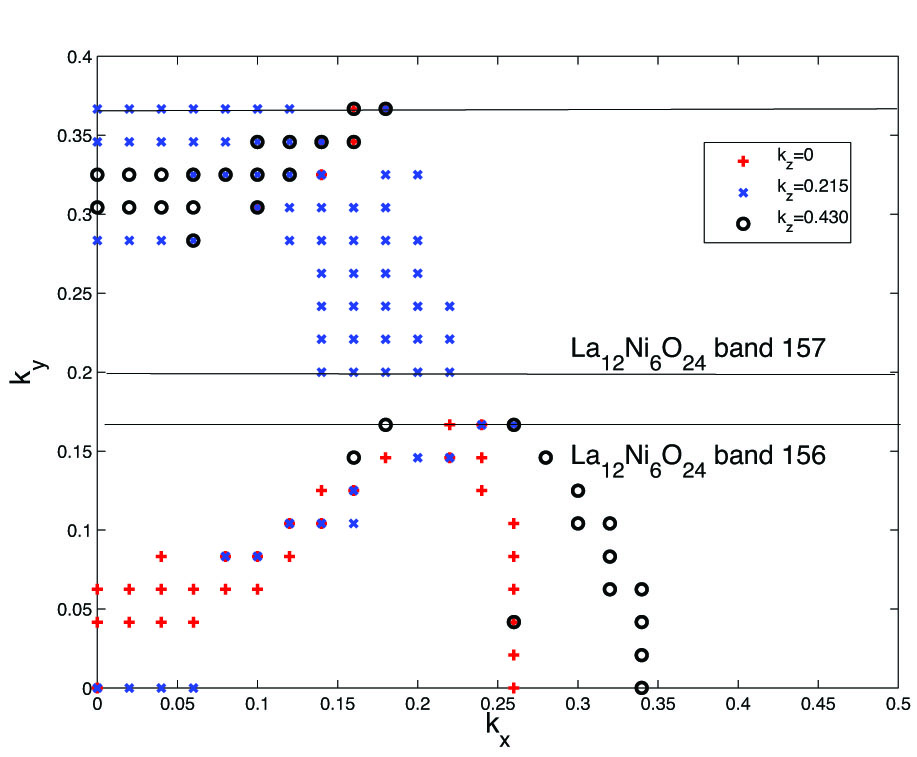}
\caption{(Color online) The FS for La$_{12}$Ni$_6$O$_{24}$.  The FS features can be understood from a folding from the FS shown in Fig.
\ref{fig6}.}
\label{fig7}
\end{figure}

\begin{figure}
\includegraphics[width=0.95\columnwidth]{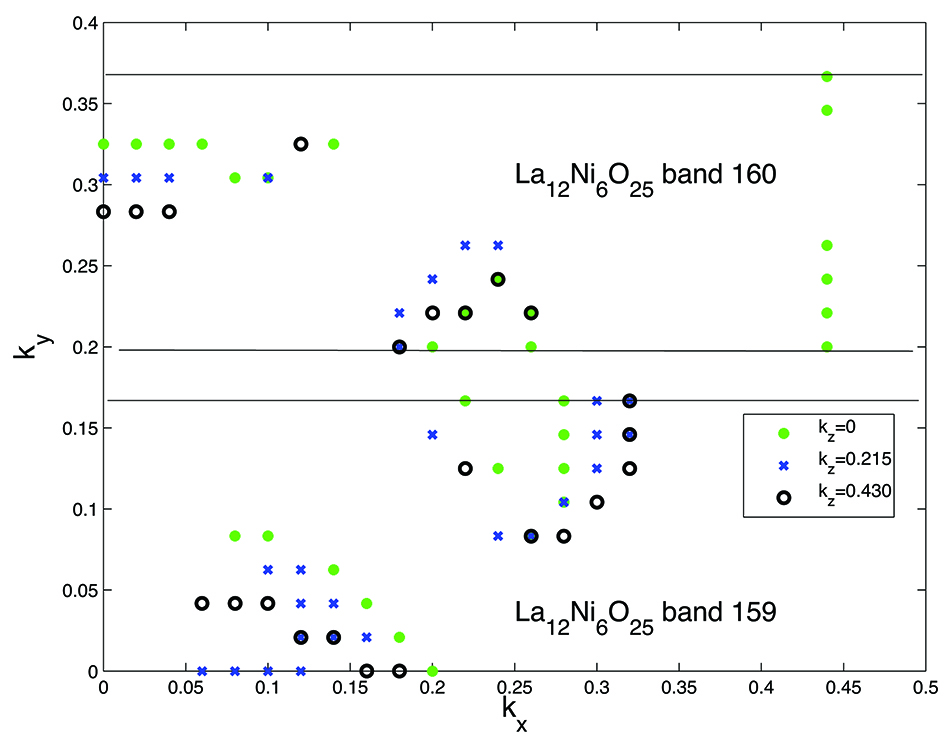}
\caption{(Color online) The FS for  La$_{12}$Ni$_6$O$_{25}$. The addition of the $O_i$ impurity leads to a fragmentation of 
the FS due to formation of
multibands.}
\label{fig8}
\end{figure}

The results for La$_2$NiO$_{4 + \delta}$ are not easy to interpret. The DOS evolves as in the cuprates, and can be
interpreted as a p-doping as $O_i$ are inserted. But the d-band is less filled compared to Cu-d, and the 
FS is quite complicated already for N=0, since $E_F$ falls
within the Ni-d bands. The FS for the double cell of undoped La$_4$Ni$_2$O$_8$ is shown in Fig. \ref{fig6}. When plotted for 
the normal simple cell it would have one $\Gamma$-centered and another piece centered in the corner of the BZ. These pieces now both appear
around the $\Gamma$ point in Fig. \ref{fig6}. The advantage of displaying the FS of the double cell is that its folding in the diagonal
direction (45 degrees with respect to the NiO bonding) easily can be compared to the FS calculated for the La$_{12}$Ni$_6$O$_{24}$
supercell, shown in Fig. \ref{fig7}. As for folding in the LCO system, there is a good agreement between folded and calculated FS's.
Next, in Fig. \ref{fig8} we show the calculated FS for La$_{12}$Ni$_6$O$_{25}$, i.e. for the case when one O$_i$ atom 
forms a kind of wire ordering in the supercell. The FS is now quite different from that of the undoped case, and as in the case
of O$_i$-wires in LCO, the FS fragmented
in small 'islands'. In contrast to LCO, it is hardly possible to see that the radii of a generic FS-piece have diminished because
of an effective doping. The local DOS near and on the O$_i$ wire is not as high as in the LCO system. The hybridization between wire states
and the NiO d-p bands is more effective, and the isolation of the wire is not as drastic as in LCO. Nevertheless, a line from a FS structure
to the right in Fig. \ref{fig8} seems to be unexplained from the ordinary bands, and could be due to the isolated O$_i$ band as in doped LCO.

The wire with new potential modulations generates small band gaps within the Ni-O bands. Original bands 
are cut into smaller 'multi-valley' bands, which for the FS leads to fragmentation. But the interpretation of the FS's for undoped
and doped LNO supercell appear not to be as simple as in LCO.

The d-band edge below $E_F$ is displaced with increasing N in a similar manner as in the cuprates,
and so the impurities are likely to create an effective p-doping within the NiO layers. An opposite and much stronger "n"-doping
would be needed to make the nickelate FS similar to the simple cuprate FS, which naively could be expected to make nickelates superconducting.
The calculated FS of the nickelate is not simple and does not resemble the FS of the cuprates, and this is independent of N.

\section {Discussion.}

In the three systems we have found evidence of electron charge transfer from the metal-oxide layers to the interstitial
$O_i$ impurities. The electron count within the atomic spheres as well as the shift of $E_F$ supports this finding in all cases.
 When the $O_i$ sites are ordered to form wires in the case of cuprates this p-type charge transfer has the same effect
on the CuO layers as the standard ways of doping, like replacement of La with Sr in LCO. The DOS becomes very high on the $O_i$ wires, but the 
latter sites are electronically isolated from the CuO layers, and the layers can maintain the generic electronic structure of common cuprates. 
Without ordering there is more overlap between not too distant $O_i$ sites, and the separation between the electronic structures
on CuO layers and $O_i$ is no longer possible. This is at least what we conclude from the absence of a typical CuO layer FS in the case
of disordered impurities. It is likely that clustering into "islands" instead of wires also can work for having both the charge transfer
and the intact FS property in the layers, but it is not known how far the charge transfer can reach between large well separated islands.  
  Thus, we suggest that a favorable effect from self-organized $O_i$ wires on $T_c$ can be caused by the doping of the CuO layers, leading
 to optimal doping with $E_F$ near the van-Hove singularity. But there is a second possibility in the case
 of wire ordering. Perpendicular periodic wires will set up a new periodicity of the potential, and this is known to make additional
 band gaps and peaks in the DOS. The strength of the potential modulation and the wavelength determines the energy and the height
 of peaks and valleys of the DOS. Other types of doping should be able to generate the same modulations, and as has been proposed
 earlier for phonon and spin waves, an optimization is required to make strong peaks to coincide with $E_F$ and the doping in order
 to expect an enhancement of $T_c$ \cite{tjapl,tj7} .  Such a mechanism should help to increase $T_c$ in any system with
 fairly simple band structures where a periodic potential can make flat bands and higher DOS at precise energies.
 Here, for cuprates with $O_i$ impurity wires one would expect  correlations between impurity concentration (through the
 distance between wires) and variations of $T_c$. Unfavorable conditions regarding
  wire separations and doping could occur and put
 $E_F$ in a valley of the DOS with low $T_c$.

 Superconducting pairing may depend on phonons, spin fluctuations or even a coupling between the two \cite{tj7}.
 Many exchange enhanced systems can be close to either a ferro magnetic (FM) or an anti-ferro magnetic (AFM) transition, and calculations
 have shown that lattice distortions may trigger the transitions. For instance, FM fluctuations in Ce and B20 type compounds like FeSi are 
 enhanced by thermal lattice disorders \cite{cevib,tjce2,tjfesi1,tjfesi2}. AFM fluctuations occur at high pressure in Fe when the lattice transforms
 from fcc to hcp structure \cite{tjfe1,tjfe2}, where they probably are important for the occurance of superconductivity. The onset of FM appears
 to play a role for anomalous thermal expansion according to calculations for compounds where the total energies are
 close for the paramagnetic and FM states \cite{invar}. Similarly, in highly doped cuprates calculations indicate that a very weak FM state
 might be preferred  over the non-magnetic one \cite{barb}. All these examples show that lattice distortions and phonons have an
 intimate connection with FM, and the calculations with selected vibrational modes indeed enforce spin waves \cite{tj7,tj1}. Here, for wire
 of $O_i$ in HBCO and LCO it is found that local moments on Cu for imposed AFM waves tend to decrease near the wire 
 (probably due to the hybridization between Cu-d and O-p on $O_i$) \cite{lco,hbco}. 
 Further works are needed for telling whether coupling between spin and phonons
 will be stronger and enhance superconductivity in the systems with $O_i$ wires, or if it will attenuate such couplings near and far
 from the wires. Ni in undoped LNO show stable FM order \cite{lno}. With the large DOS near the $O_i$ wire one could expect enforcement
 of the local Ni moments near the wire. But the calculation show that the local DOS on Ni near the wire is only about 2/3 of
 the Ni DOS in undoped LNO and 1/2 of the Ni far from the wire. The FM calculations reflect these DOS variations, since the moment on Ni
 is highest far from the wire (0.25 $\mu_B$), lower in pure LNO (0.18 $\mu_B$), while extremely small close to $O_i$.
 This shows that the large DOS on (ordered) $O_i$ is very localized to the wire itself (dominated by O-p), but the proximity to the
 nearest Ni makes an essential non-magnetic layer of Ni. If FM was the only reason for the absens of superconductivity in LNO, then
 one could hope for pairing within a thin LNO layer near $O_i$ wires. However, as mentioned above, the FS of LNO is very complex and has
 no resemblance with that of the cuprates, and it is likely that a simple FS is one of the conditions for pairing in the metal-oxide
 layers. 


\section{Conclusion.}
The main conclusion is that interstitial $O_i$ impurities lead to p-doping of the metal-oxide planes in all of the studied
systems. Further, ordering of the impurities into wires makes an electronic isolation of the wires and the layers in the cuprates,
so that the CuO layers behave much like ordinary p-doped cuprates. 
The wires of O$_i$ impurities in may contribute to a potential modulation with the formation
 of a new quasi 1D band.
The calculated very large DOS peak at $E_F$ is not made from such an effect, but is caused by localized states on the wire.
The positive effect on $T_c$ from clustering of O impurities might originate from the hole
doping that the excess O's provide to the spacial regions far from the dopants. The latter
and the wire regions are spatially and electronically separated, so that
superconductivity can take place without negative disturbance from the impurities.

These findings are useful for future experiments, both for spectroscopy and for searching correlation between $T_c$ variations and
interstitial doping of oxygen. In fact, theory predicts an oscillatory behavior of DOS height at $E_F$ when impurities cause small
periodic perturbations of the potential, but this has not been verified experimentally. Direct measurements of $T_c$ variations in cuprates
should constitute a sensitive test of such correlations. The nickelate system has a much more complicated FS and higher metallic DOS near $E_F$
than the cuprates, also after doping. There are no indications that LNO can be doped and develop its band structure into something
like that of cuprate superconductors.

In conclusion the present work opens a new scenario for the mechanism of high T$_c$ superconductivity in cuprates. 
We provide evidence that ubiquitus oxygen wires of finite size, formed in oxygen doped cuprates, give
in their proximity, the first set of states with broken Fermi surfaces forming quasi 1D electronic states discussed here and 
far away from the wires, a second set of states made of a 2D strongly correlated doped Mott insulator.
The  wires formation gives first quantum confined localized states near the wires 
which coexist with second delocalized states in the Fermi-surface (FS) of doped cuprates.
In the novel scenario  for high T$_c$ superconductivity proposed by this work the first set of electronic states is identified 
as an array of Kitaev wires\cite{k1} giving Majorana bound states \cite{k2} 
which are proximity-coupled to the second set of states, far from the oxygen wires, realizing the 2D d-wave superconductor \cite{k3,k4}


\end{document}